\documentclass[12pt]{iopart}
\usepackage{iopams}
\usepackage{graphicx}

\begin{document}

\title{Trapped-ion probing of light-induced charging effects on dielectrics}
\author{M.~Harlander$^{1}$, M.~Brownnutt$^1$, W.~H\"{a}nsel$^{1,2}$ and R.~Blatt$^{1,2}$ }

\address{$^1$ Institut f\"{u}r Experimentalphysik,
Universit\"{a}t Innsbruck,
Technikerstrasse 25,
A-6020 Innsbruck, Austria}

\address{$^2$ Institut f\"{u}r Quantenoptik und Quanteninformation
der \"{O}sterreichischen Akademie der Wissenschaften,
Technikerstrasse 21a,
A-6020 Innsbruck, Austria}

\ead{max.harlander@uibk.ac.at}

\date{\today}

\begin{abstract}
We use a string of confined $^{40}$Ca$^+$ ions to measure perturbations to a trapping potential which are caused by light-induced charging of an anti-reflection coated window and of insulating patches on the ion-trap electrodes.
The electric fields induced at the ions' position are characterised as a function of distance to the dielectric, and as a function of the incident optical power and wavelength.
The measurement of the ion-string position is sensitive to as few as $40$ elementary charges per $\sqrt{ \mathrm{Hz}}$ on the dielectric at distances of order millimetres, and perturbations are observed for illumination with light of wavelengths as long as 729\,nm.
This has important implications for the future of miniaturised ion-trap experiments, notably with regards to the choice of electrode material, and the optics that must be integrated in the vicinity of the ion.
The method presented can be readily applied to the investigation of charging effects beyond the context of ion trap experiments.
\end{abstract}

\maketitle

\section{Introduction}

As segmented ion traps are scaled to ever smaller dimensions, and hold ever more ions \cite{Kielpinski:2002}, questions necessarily arise about the size and location of the experimental apparatus accompanying the trap \cite{Steane:2007, Kim:2009}.
To collect as much light as possible there are clear advantages in having optics very close to the ion.
This requires either a vacuum window in close proximity to the trap or optics within the vacuum, such as lenses \cite{Streed:2008}, fibres \cite{Kim:2009}, or cavity mirrors \cite{Keller:2004, Russo:2009, Herskind:2009}.
While experiments currently operate with ion-dielectric separations of a few millimetres to centimetres, proposals exist for reducing this to a few hundred microns to facilitate, for example, strong ion-cavity coupling \cite{Keller:2004}, or coupling ions to nanomechanical resonators \cite{Hensinger:2005}.

Insulating materials can accumulate charge which is hard to dissipate and drifts slowly over time.
Such materials in or near ion traps can thus frustrate efforts to obtain a well-controlled trapping potential.
Slowly drifting charges in the vicinity of trapped ions can give rise to problems of excess micromotion \cite{Berkeland:1998} and difficulties in reliably addressing ions with focussed laser beams.
A common approach is therefore to avoid insulating materials in the vicinity of the ion as much as possible.
It is clear, however, that the pragmatic solution of simply keeping insulators far removed becomes increasingly untenable as questions of trap miniaturisation move to the fore.

In addition to miniaturisation of optical components around the trap, there are significant advantages to miniaturising the trap itself.
Smaller traps allow for higher ion motional frequencies, facilitating faster gate speeds and simplifying fast ion transport.
However, as ion-electrode separations are reduced, ions become increasingly sensitive to fluctuating charges on the electrodes' surfaces \cite{Turchette:2000}.
Thermally induced fluctuations are believed to play an important role in the anomalous heating observed in ion traps \cite{Turchette:2000, Camp:1991}.
While such effects can be mitigated by cooling traps to cryogenic temperatures \cite{Labaziewicz:2008} this provides, at best, only an engineering solution to a little-understood physics problem.
Possibly-related phenomena are held responsible for decoherence in super-conducting qubit experiments \cite{Oh:2006, Schriefl:2006}.
Here the coherence is destroyed by electric field modulation due to thermally activated dipoles, so-called ``two-level fluctuators''.
Ideally, the source and nature of these fluctuations should be understood.

Furthermore, issues regarding surface charging are of interest in the context of neutral atom experiments.
When neutral atoms are brought close enough to dielectric or metallic surfaces they become sensitive to the associated local electric fields \cite{Lin:2004, Obrecht:2007, Hunger:2010}.
Understanding the charging effects in materials used for such experiments is of great import for correctly interpreting the results.
In particular, strong light fields may impinge on materials being investigated, such as in evanescent-field experiments \cite{Westbrook:1998, Bender:2010}, and may potentially charge the surfaces.

In this article a systematic and quantitative investigation of charging effects on an insulator is undertaken, using trapped ions as a highly sensitive probe for the electric field.
Specifically, charging effects due to optical illumination of a dielectrically-coated glass plate placed a few millimetres from an ion chain are characterised as a function of the ion-glass separation.
The open geometry of the planar ion trap used is advantageous as it grants easy access to the ions for the material under investigation.
Additionally, the charging of insulating patches on the copper trap electrodes is observed and investigated.
These patches may be a complete insulating layer, and are presumably formed from an oxide.
Charging of the glass plate and of the electrodes is characterised as a function of the incident laser wavelength, in the range 375\,nm - 729\,nm.

Ions held in a radio frequency (RF) trap are known to be sensitive field probes: a single ion has previously been used to map out the electromagnetic field of an optical resonator \cite{Guthohrlein:2001}, and fluctuating electric fields have been measured by motional heating of trapped ions \cite{Labaziewicz:2008, Deslauriers:2006a}.
It has recently been proposed that an ion in a ``stylus trap" could be used to sensitively measure electric and magnetic fields \cite{Maiwald:2009}.
In the work presented here, up to four $^{40}$Ca$^+$ ions are trapped in a standard surface ion trap, and a glass plate with a dielectric anti-reflective coating can be moved relative to the ions using a mechanical vacuum feedthrough.
By observing the change in position of a three-ion string, the presence of $\sim 40$ elementary charges can be detected at a distance of 1.2\,mm within one second.

\section{Apparatus and Method}

\begin{figure}
\includegraphics[width=15cm,keepaspectratio]{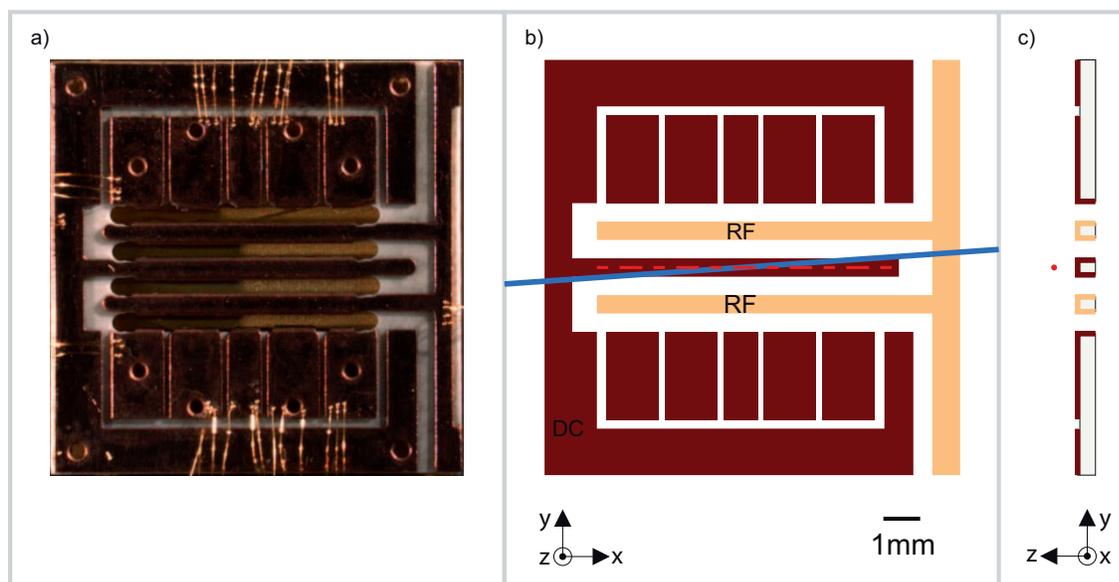}
\caption{\label{fig:TrapLayout}
Ions are trapped using a Cu-on-PCB planar ion trap
\cite{Splatt:2009} (a).
The geometry is shown schematically in plan view (b) and cross section (c).
An RF voltage of $\sim 340$\,V amplitude is applied to the RF electrodes to provide radial confinement, and DC voltages of up to 50\,V are applied to the remaining electrodes for axial confinement.
The red dashed line (b)/dot (c) gives the position of the RF null along which the ions are confined, $\sim$800\,$\mu$m above the surface.
The ions are cooled by lasers, indicated by the blue line in (b), parallel to the trap surface, and at 4$^{\circ}$ to the trap axis.}
\end{figure}

\begin{figure}
\includegraphics[width=10cm]{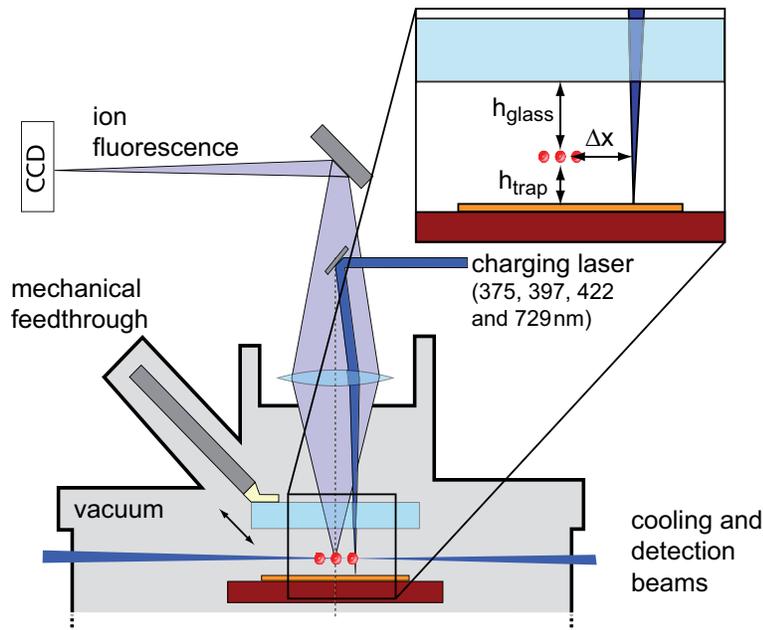}
\caption{\label{fig:OpticalSetUp}
Optical setup.
A glass plate attached to a mechanical vacuum feed through can be held at different distances (1\,mm$\lesssim d \lesssim$10\,cm) to the ion string (not to scale).
Via a small mirror in the path of the fluorescence light, the glass plate and the central trap electrode can be illuminated with laser light of different wavelengths.}
\end{figure}

All charge sensing experiments described here are performed with a string of ions in a five-wire planar trap \cite{Leibrandt:2007, Splatt:2009}.
The trap, depicted in figure~\ref{fig:TrapLayout}, is made from copper electrodes on a vacuum-compatible printed-circuit-board (PCB) material (Rogers 4350), and mechanical cut-outs reduce the amount of exposed dielectric surface (courtesy of the group of I.~L.~Chuang, MIT).
A string of $^{40}$Ca$^+$ ions is trapped $\sim$800\,$\mu$m above the surface (the exact height of the RF null is dependent on the position of the glass plate, shown in figure~\ref{fig:OpticalSetUp}).
In the experiments described, the ion string typically has motional frequencies of 2$\pi\times$(90, 230, 790)\,kHz in the axial and two radial directions respectively.
The ions are laser-cooled on the S$_{1/2}$-P$_{1/2}$ transition using 397\,nm light \cite{Schmidt-Kaler:2003a} so that they form a stable ion crystal, where each ion can be individually resolved.
The cooling beam runs parallel to the surface of the trap, at a slight angle to the trap axis (figure~\ref{fig:TrapLayout}b).

As depicted in figure~\ref{fig:OpticalSetUp} the fluorescence light is imaged on an EM-CDD camera.
A glass plate with an anti-reflective coating~\footnote{Antireflective coating from Laseroptik Garbsen} can be brought close ($\sim$1\,mm) to the ions using a mechanical vacuum feedthrough, or withdrawn completely to be $\sim$10\,cm distant, out of the optical path.
Using a small mirror in the path of the fluorescence light, different charging laser beams can be shone through the glass plate and onto the trap.
The beams are focussed to a diameter of below $100\,\mathrm{\mu m}$ and the beam pointing is calibrated by observing the position of the laser beam on the trap with the CCD camera.
Once this position is known, narrow-band dielectric filters are added in front of the camera for transmission at 397\,nm, to allow detection of light scattered by the ion despite the presence of stray light from the trap surface.
A series of CCD images is taken, each comprising 10 accumulations of $100\,\mathrm{ms}$ exposure, and $1.9\,\mathrm{\mu m}$ resolution to measure the positions of the ions.
By fitting the individual ion positions and averaging over all ions, the precision of the string's centre of mass position is increased beyond the resolution of the camera picture.
The axial motional frequency of the ions gives information regarding the curvature of the confining potential.
This in turn allows changes in the ions' position to be used to calculate the axial forces acting on the ion string, and so infer the axial electric field at the ions' position.
For the study of charging by 397\,nm light the discrimination between ion fluorescence and stray light is impossible using optical filters.
Instead, a pulsed scheme with shorter charging durations and interleaved detection phases is applied.

\section{Results}

\begin{figure}
\includegraphics[width=15cm]{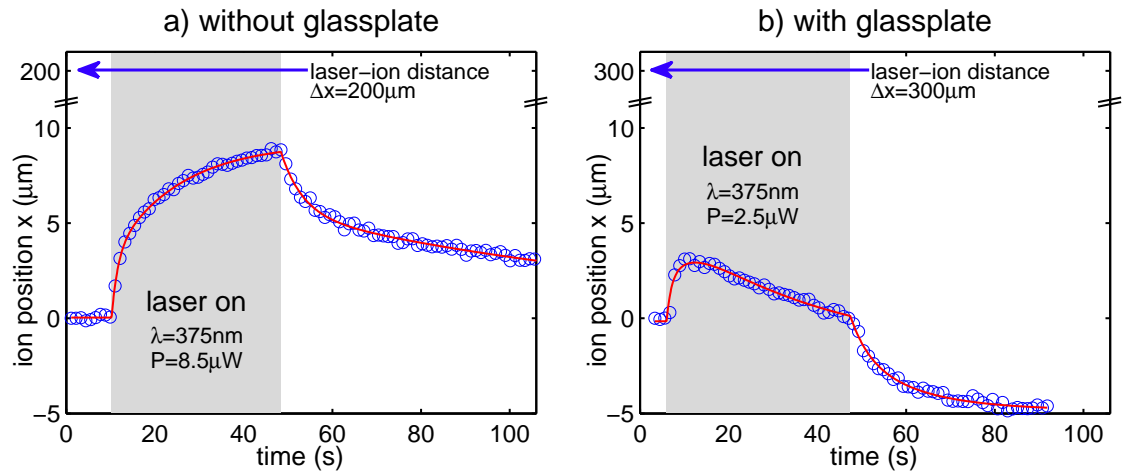}
\caption{\label{fig:TypicalDrift} Typical example of the axial positions of the ion string as the illuminating laser is turned on (grey-shaded area) and off.
(a) No glass plate is present, so the charging leading to the ions' displacement towards the laser beam occurs at the trap surface only.
The data was taken with $8.5\,\mathrm{\mu W}$ of $\lambda=375\,\mathrm{nm}$ light, focussed on the trap at $\Delta x = 200\,\mathrm{\mu m}$ away from the ion.
In figure (b) the ions' movement towards the laser beam is superimposed with a repulsive effect, that arises at the glass plate.
The data was taken with $2.5\,\mathrm{\mu W}$ of $\lambda=375\,\mathrm{nm}$ light, focussed on the trap at $\Delta x = 300\,\mathrm{\mu m}$ away from the ion, with the glass plate a distance of $h_{\rm glass} \approx 2\,$mm from the ion.
The solid lines are the exponential fits to the data.}
\end{figure}

Two different effects are observed, depending on the presence or absence of the glass plate.
Throughout the following section, graphics on the left-hand side refer to the situation without the glass plate, while the graphics on the right describe results obtained in the presence of the glass plate.
Figure~\ref{fig:TypicalDrift} shows the centre of mass position of the ion string as a function of time during and after laser illumination of the centre electrode.
Here, a laser beam ($\lambda = 375\,\mathrm{nm}$) is incident to the right of the ion string at $\Delta x=200\mathrm{\mu m}$.
In the first case (a) an attractive force acts on the ions as the substrate is illuminated (grey-shaded area), making them move at $\sim$1$\,\mathrm{\mu}$m/s towards the laser beam.
This movement saturates on a timescale of $\sim$1$\,$min.
Switching off the laser beam causes the ion string to relax to close to its original position on a somewhat longer timescale.
In the second case (b), with the glass plate placed close to the ions, the attractive force during laser illumination is superimposed with a repulsive one, which saturates dependent on the applied laser intensity on a timescale of minutes.
The displacement caused by this additional force subsequently remains unchanged for days.
Additionally, when the glass plate is present, room-light from fluorescent tubes affects the position of the ion string. Light from Tungsten filament lamps has no observable effect.

The observed results point to two different charging effects: (a) negative charging of the copper electrode surface and (b) positive charging of the glass substrate.
While the exact nature of the copper charging is not known, it is assumed that the surface is partially or completely covered by a thin insulating layer, probably an oxide \cite{Kim:2010}.
One might expect that charges directly above the copper surface produce negligible fields, because image charges form directly beneath them and shield their effect.
However, the resulting field is that of a dipole, which drops off more quickly than a monopole, but which is still easily detected by the sensitive ion probe.

\begin{figure}
\includegraphics[width=15cm]{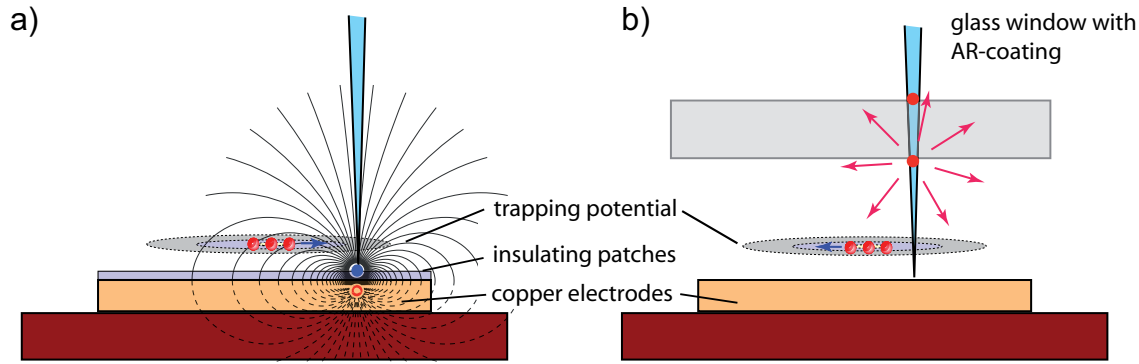}
\caption{\label{fig:ChargingEffects}
Two distinct charging effects are observed.
The first (a) is consistent with electrons being held on a thin oxide layer above the copper.
The electrons and their image charges create a dipole which attracts the ion string.
The second effect (b) is consistent with electrons being ejected from the front and back surfaces of the glass plate leaving two regions of positive charge which repel the ion string.}
\end{figure}

Analyzing the data more quantitatively leads to the following model for the charging processes: under illumination with laser light local charges are produced with a wavelength-dependent production rate, $P$.
It has been experimentally verified that this production rate scales linearly with incident laser power for both charging processes.
The presence of existing charges creates a barrier inhibiting the creation of new charges.
This may, to first order, be accounted for by considering a production rate $P=P_0(1-\delta\cdot n_q)$, where $n_q$ denotes the number of charges present and $\delta$ describes the barrier effect.
In addition, produced charges may be neutralised at a rate $\gamma$.
The total process is then described by the differential equation
\begin{equation}
\dot{n}_{q}= P_0(1-\delta \cdot n_{q}) -\gamma \cdot n_q \,
\end{equation}
where the values for $P_0$, $\delta$, and $\gamma$ depend on the process and the (local) material properties. The deduced time dependence of the number of charges present during and after laser illumination then becomes
\begin{equation}
\label{eq:ChargingSolution}
n_q = \left\{ \begin{array}{ll}n_{\rm eq}\cdot[1-\exp(\gamma_{\rm on} t)] \\ n_0\cdot \exp(-\gamma_{\rm off} t) \end{array}\right.
\end{equation}
where the inverse settling times are $\gamma_{\rm on}=\gamma+P_0\,\delta$ and $\gamma_{\rm off}=\gamma$;  $n_{\rm eq}=P_0/\gamma_{\rm on}$ is the steady state solution under illumination,  and a constant, $n_0$, reflects the amount of charge present at the moment the laser is switched off. These two solutions represent the core for fitting both charging effects where, for the charging of the glass plate, the absence of relaxation is accounted for by setting $\gamma=0$.

Charging of the copper electrode may be explained by the excitation of electrons onto insulating patches directly above the bulk material.
Together with the induced image charge this creates a small electric dipole.
Such an excitation represents a metastable state which may easily relax back to a neutral state ($\gamma\neq0$).
This hypothesis is corroborated by the investigation of the axial electric field.
The initial velocity of the ion string during laser illumination is proportional to the electric field per charge multiplied by the charge production rate.
By changing the position of laser illumination with respect to the ions' position, the lateral field per created charge is mapped out.

\begin{figure}
\includegraphics[width=15cm]{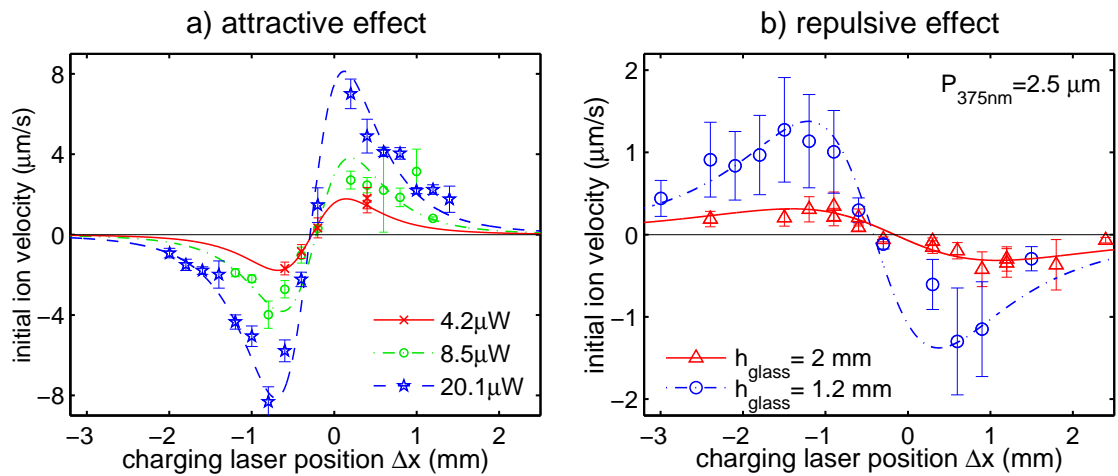}
\caption{\label{fig:RateVsLasPos}
Initial ion velocity as a function of laser beam position ($\lambda = 375\,\mathrm{nm}$) due to charging of (a) the trap surface for different laser powers and (b) the glass surface for two ion-glass distances.
The lines are fits according to the proposed model.}
\end{figure}

Figure~\ref{fig:RateVsLasPos}a shows the initial velocities for different powers of laser illumination for $\lambda = 375\,\mathrm{nm}$.
The behaviour for other wavelengths is similar.
The fitted curves correspond to the velocity expected from the axial component of a dipole field as described by
\begin{equation}
\dot{x}_{\rm ion}=\dot{D}_{\rm rel}\cdot\frac{h_{\rm trap}\cdot{\Delta x}}{\left( \Delta x^2 + h_{\rm trap}^2 \right)^{5/2}} \,.
\end{equation}
Here $\dot{D}_{rel}=\frac{3 q_{\rm Ca}}{4 \pi \epsilon_0 m_{\rm Ca} \omega^2}\cdot \dot{n}_{q} q_{e} r_{\rm dip}$ is the fit parameter for the net production rate of surface charges (hence dipoles), with $q_{\rm Ca}$ and $m_{\rm Ca}$ the charge and mass of a calcium ion, $\omega$ the axial trap frequency, $q_e$ the electron charge and $r_{\rm dip}$ the distance between charge and image charge, which is twice the thickness of the insulating layer.
The symbols $h_{\rm trap}$ and $\Delta x$ denote the ion-electrode distance and the axial distance between ion and laser beam respectively.
The measured data excludes any significant contribution from monopole or quadrupole (and higher order) terms.

As the thickness of the insulation patches, and hence the extension of the dipole $r_{\rm dip}$, is unknown the quantum efficiency of the dipole production process, $\eta_{\mathrm{dip}}$, can only be expressed including this unknown value.
For the wavelengths $\lambda = (375, 422, 729)\,\mathrm{nm}$ we deduce $\eta_{\mathrm{dip}}= (14, 62, 0.2)\cdot10^{-9}/(r_{\mathrm{dip}}/\mathrm{\mu m})$ respectively.
The data does not provide sufficient information to propose a model that describes the wavelength dependence of the dipole-production rates, e.g. the significant reduction of the rate at $375\,\mathrm{nm}$ compared to that at $422\,\mathrm{nm}$.

It is interesting to note that the time dependence of the ion-string position for electrode charging is not well fitted with equation~(\ref{eq:ChargingSolution}) alone.
This is attributed to the fact that there is a multitude of metastable states to which the electrons may be excited, given the likely inhomogeneous nature of the electrodes' surface.
Each state may have its own excitation probability and decay constant, leading to an inhomogeneous broadening.
Heuristically, we take this into account by expressing the total number of charges as the result of two separate charging processes with independent production rates and relaxation constants.
The individual constants do not have a direct physical meaning and may depend on the specific investigated time span.
However, as an indication, the derived relaxation constants for the charge production (laser on) are $1/\gamma_{\mathrm{on,1}}=2\,\mathrm{s}$ and $1/\gamma_{\mathrm{on,2}}=16\,\mathrm{s}$ and for the decay (laser off) $1/\gamma_{\mathrm{off,1}}=5\,\mathrm{s}$ and $1/\gamma_{\mathrm{off,2}}=120\,\mathrm{s}$.
The limitation to two production processes does not affect the evaluation of the above quantum efficiencies since they only depend on the total initial charge production rates.

When the dielectric glass plate is brought close to the surface trap it might be expected that electrons are extracted from the trap electrodes by light or by the trap voltage and settle on the glass.
While we cannot exclude that this negative charging effect occurs to a small degree, the only detectable effect is a net positive charging at the position where the laser penetrates the glass.
The induced repulsive effect on the ion string can be separated from the surface effects because the surface effects relax on a small time scale while the glass charging remains.
We attribute the positive charging to the ejection of electrons from the antireflective coating of the substrate.
Since these holes are produced in a dielectric surrounding, no electrons can be reabsorbed and therefore the charge creation is irreversible.
In figure~\ref{fig:TypicalDrift}b the ions' movement is fitted with the model which sums both the attractive and repulsive effects.
The ion velocities induced by the glass charging are shown in figure~\ref{fig:RateVsLasPos}b for two different ion-glass distances $h_{\mathrm{glass}}$ at a laser power of $\sim2.5\,\mathrm{\mu W}$.
The charging saturates on a timescale of $1/\gamma_{\mathrm{on}} = 38\,\mathrm{s}$.
The fitted curves indicate the expected axial field for a point charge located at the glass surface nearest to the ion trap.
The charging effect at the backside of the glass plate has been neglected as the thickness of the glass (5\,mm) is large compared to the ion-glass distance.
In addition to this monopole-like contribution
\begin{equation}
\dot{x}_{\rm ion}=\dot{Q}_{\rm rel} \cdot \frac{\Delta x}{\left( \Delta x^2 + h_{\rm glass}^2 \right)^{3/2}}
\label{eq:Monopole}
\end{equation}
the small shielding due to an image charge with opposite sign and appropriate distance (at $h_{\rm glass} + h_{\rm trap}$ below the trap surface) has been taken into account.
No significant contribution to the field from dipoles on the glass surface is observed.
In equation~(\ref{eq:Monopole}) $\dot{Q}_{\rm rel}=\frac{q_{\rm Ca}}{4 \pi \epsilon_0 m_{\rm Ca} \omega^2} \cdot \dot{n}_{q}\, q_{\rm e}$ is the relative charge production rate, $h_{\rm glass}$ is the ion-glass distance and $n_q$ the number of positive charges created on the glass plate.
With the known trap frequency $\omega \sim \, 90\mathrm{kHz}$ the charge-production efficiency is calculated to be $\eta_{375} = 1.2\cdot10^{-10}$ electrons per photon at $\lambda = 375\,\mathrm{nm}$ and $\eta_{397}=0.4 \cdot 10^{-10}$ at $\lambda = 397\,\mathrm{nm}$.
For $\lambda = 422\,\mathrm{nm}$ and $\lambda = 729\,\mathrm{nm}$ no movement of the ion string, and therefore charging of the glass, could be detected at the given level of sensitivity.

\section{Estimation of the charge sensitivity}

Using the imaging method described here with a three-ion crystal, an upper limit for the position uncertainty is computed from the residuals to be $\delta$x = 0.12\,$\mu$m for one second acquisition time.
The uncertainty in position scales with the inverse square root of the averaging time and of the number of ions.
The time and ion-number dependence can thus be removed from the uncertainty to yield a sensitivity of $0.12\,\mu\mathrm{m}\cdot\sqrt{3/\mathrm{Hz}}=0.21\,\mu\mathrm{m}/\sqrt{\mathrm{Hz}}$.
Given the trap frequency of 90\,kHz, this corresponds to an electric field sensitivity (for a single calcium ion) of $30\,\mathrm{mV} / ( \mathrm{m}\cdot\sqrt{\mathrm{Hz}}$), and a force sensitivity of $4.5\,\mathrm{zN} / \sqrt{\mathrm{Hz}}\; (\mathrm{zN} = 10^{-21}\,\mathrm{N})$.
With the laser displacement, $\Delta x$, at the most sensitive position (peaks of the curves in figure~\ref{fig:RateVsLasPos}) and with a glass-ion distance of 1.2\,mm, this is equivalent to detecting the presence of 40 elementary charges with three ions within one second.
For similar charge sensitivities, near-field probes such as scanning atomic force microscopes are usually required \cite{Binning:1985}.
However, using the ion trap as a measuring device has the advantage that the sensor leaves the surface completely unobstructed and does not interfere with the measured effect.

Very recently, measurements as sensitive as $0.4 \, \mathrm{zN} / \sqrt{\mathrm{Hz}}$ have been reported in a Penning-trap, using RF excitation of a driven harmonic system near resonance \cite{Biercuk:2010}.
For comparison, this is equivalent to a single-ion field sensitivity of $0.3 \, \mathrm{mV} / (\mathrm{m} \cdot \sqrt{\mathrm{Hz}} )$ and a single-ion force sensitivity of $ 0.05\,\mathrm{zN} / \sqrt{\mathrm{Hz}}$.
This system measures a pulsed, radio-frequency force, and infers the magnitude from Doppler velocimetry.
This technique is applicable to the ``high frequency'' range over which the trap frequency can be tuned.
By contrast, while being a factor of 100 less sensitive, the simple technique described here measures slowly varying forces in real time.

Provided the precision of the ion-position measurement is limited by the imaging technique, rather than the ion's thermal spread, the measurement sensitivity scales with the inverse square of the trap frequency.
Beyond this limit, it scales with the inverse of the trap frequency.
It would thus be relatively simple to improve the sensitivity of the ion probe method by an order of magnitude by reducing the ions' motional frequency.
The sensitivity could be further increased using homodyne detection (e.g. observing micromotion sidebands for transverse excursion) or using quantum detection schemes \cite{Maiwald:2009}.

\section{Summary \& Outlook}
While there has previously been a number of speculative assumptions regarding the perturbing effects of insulators in the vicinity of trapped ions, the true nature of the perturbations was untested.
Here, two distinct mechanisms for charging dielectrics are revealed, using the trapped ions themselves as sensitive and non-invasive field probes.
Laser or room light incident on an optically-coated glass surface was observed to locally create positive charges, while laser light impinging on the copper trap electrodes caused the accumulation of negative charges on the electrode surface, presumed to be located on insulating patches.
The low activation energy for the latter process of less than 1.6\,eV suggests the direct light-induced occupation of these patches.
Within the resolution of the measurement, no spatial variation was observed in the patch distribution.
These findings are of interest for a wide variety of experiments that involve probing of particles (neutral or charged) in close proximity to surfaces.
For ions this effect is sufficiently strong that the operation of ion traps may be compromised even when the distance to the charged material is on the order of a centimetre.
This is of particular importance in the field of miniaturised ion traps, where the integration of optical high-finesse microcavities with ion traps necessitates the placement of optically coated glass surfaces in close proximity to the ions.
This work may also contribute to the discussion about patch-charge fluctuations that are thought to be responsible for heating of ions in proximity to conductive surfaces.
With a tightly focussed laser beam it may be possible to directly resolve single patches of sizes on the order of a micrometre.
Finally, this method provides a general tool to study local charging of any surface that can be brought close to an ion trap.

\ack
We thankfully acknowledge that the ion trap used is a courtesy of the group of I.~L. Chuang, MIT.
We also gratefully acknowledge the support of the Austrian Science Fund (FWF), the EU network SCALA, and the EU STREP project MICROTRAP,
and of the Institut f\"{u}r Quanteninformationsverarbeitung.

\section*{References}

\providecommand{\newblock}{}

\end{document}